\documentclass[preprint,aps]{revtex4}
\usepackage{mathrsfs}
\usepackage{graphicx}
\usepackage{multirow}
\usepackage{makecell}
\usepackage{booktabs}
\usepackage{bm}

\begin{document}

\title{Evidence for an Additional Symmetry Breaking from Direct Observation of Band Splitting in the Nematic State of FeSe Superconductor}

\author{Cong Li$^{1,2\sharp}$, Xianxin Wu$^{1,3\sharp}$, Le Wang$^{1,2\sharp}$, Defa Liu$^{1,4}$, Yongqing Cai$^{1,2}$, Yang Wang$^{1,2}$, Qiang Gao$^{1,2}$, Chunyao Song$^{1,2}$, Jianwei Huang$^{1,2}$, Chenxiao Dong$^{1,2}$, Jing Liu$^{1,2}$, Ping Ai$^{1,2}$, Hailan Luo$^{1,2}$, ChaoHui Yin$^{1,2}$, Guodong Liu$^{1}$, Yuan Huang$^{1}$, Qingyan Wang$^{1}$, Xiaowen Jia$^{5}$, Fengfeng Zhang$^{6}$, Shenjin Zhang$^{6}$, Feng Yang$^{6}$, Zhimin Wang$^{6}$, Qinjun Peng$^{6}$, Zuyan Xu$^{6}$, Youguo Shi$^{1,2}$, Jiangping Hu$^{1,2,9}$, Tao Xiang$^{1,2,9}$, Lin Zhao$^{1,*}$ and X. J. Zhou$^{1,2,7,8,*}$
}

\affiliation{
\\$^{1}$Beijing National Laboratory for Condensed Matter Physics, Institute of Physics, Chinese Academy of Sciences, Beijing 100190, China
\\$^{2}$University of Chinese Academy of Sciences, Beijing 100049, China
\\$^{3}$Institute for Theoretical Physics and Astrophysics, Julius-Maximilians University of Wurzburg, Am Hubland, D-97074 Wurzburg, Germany
\\$^{4}$Max Planck Institute of Microstructure Physics, Weinberg 2, Halle 06120, Germany
\\$^{5}$Military Transportation University, Tianjin 300161, China
\\$^{6}$Technical Institute of Physics and Chemistry, Chinese Academy of Sciences, Beijing 100190, China
\\$^{7}$Songshan Lake Materials Laboratory, Dongguan 523808, China
\\$^{8}$Beijing Academy of Quantum Information Sciences, Beijing 100193, China
\\$^{9}$Collaborative Innovation Center of Quantum Matter, Beijing 100190, China
\\$^{\sharp}$These people contributed equally to the present work.
\\$^{*}$Corresponding authors: lzhao@iphy.ac.cn and XJZhou@iphy.ac.cn
}

\date{November. 19, 2019}


%\begin{abstract}

%\end{abstract}

\maketitle

%%Abstract

{\bf The iron-based superconductor FeSe has attracted much recent attention because of its simple crystal structure, distinct electronic structure and rich physics exhibited by itself and its derivatives. Determination of its intrinsic electronic structure is crucial to understand its physical properties and superconductivity mechanism. Both theoretical and experimental studies so far have provided a picture that FeSe consists of one hole-like Fermi surface around the Brillouin zone center in its nematic state. Here we report direct observation of two hole-like Fermi surface sheets around the Brillouin zone center, and the splitting of the associated bands,  in the nematic state of FeSe by taking high resolution laser-based angle-resolved photoemission measurements. These results indicate that, in addition to nematic order and spin-orbit coupling, there is an additional order in FeSe that breaks either inversion or time reversal symmetries.  The new Fermi surface topology asks for reexamination of the existing theoretical and experimental understanding of FeSe and stimulates further efforts to identify the origin of the hidden order in its nematic state.}

%%Introduction

In iron-based superconductors, FeSe has the simplest crystal structure which consists solely of the basic building block, the FeSe layers, that dictates the superconductivity\cite{Hsu_PNAS}. It undergoes a structural phase transition at around 90 K to enter a nematic state but without a long range magnetic order\cite{McQueen_PRL}.  Superconductivity of FeSe can be significantly enhanced under high pressure\cite{SMedvedev_NM,JPSun_NC}. It represents a unique system that superconductivity and nematicity coexist in the superconducting state,  giving rise to its distinct electronic structure and superconducting gap structure\cite{JMaletz_PRB,KNakayama_PRL,TShimojima_PRB,Watson_PRB,Watson2_PRB,Watson_NJP,Watson_JPSJ,PZhang_PRB,Suzuki_PRB,Xu_PRL,DFLiu_PRX2018,Rhodes_PRB,THashimoto_NC,Kushnirenko_PRB,Pfau_PRL,Kasahara_PNAS,Sprau_Science,Hanaguri_PRL,Terashima_PRB,Watson_PRL}. Many new superconductors have been derived from FeSe with enhanced superconductivity, including A$_x$Fe$_{2-y}$Se$_2$ (A=K, Rb, Tl and etc.)\cite{JGGang_PRB,TPYing_JACS} and other organic intercalated superconductors\cite{TPYing_SciRep,Maziopa_JPCM,Lucas_NM,Hatakeda_JPSJ,MZShi_PRM}, (Li,Fe)OHFeSe\cite{Lu_NM}, heavily electron-doped FeSe through gating or potassium deposition\cite{BLei_PRL2016,BLei_PRB2017,JJSeo_NC}, and in particular, single-layer FeSe/SrTiO$_3$ films with a record high $T_{c}$\cite{QYWang_CPL,DFLiu_NC,SLHe_NM,SYTan_NM,JJLee_Nature}. Determination of the electronic structure is essential to understanding the physical properties and superconductivity mechanism in bulk FeSe and its many derivatives. So far, all the angle-resolved photoemission (ARPES) measurements report only one hole-like Fermi surface around the Brillouin zone (BZ) center in its nematic state\cite{JMaletz_PRB,KNakayama_PRL,TShimojima_PRB,Watson_PRB,Watson2_PRB,Watson_NJP,Watson_JPSJ,PZhang_PRB,Suzuki_PRB,Xu_PRL,DFLiu_PRX2018,Rhodes_PRB,THashimoto_NC,Kushnirenko_PRB,Pfau_PRL}. Scanning tunneling microscope (STM) results\cite{Kasahara_PNAS,Sprau_Science,Hanaguri_PRL} and quantum oscillation measurements\cite{Terashima_PRB,Watson_PRB,Watson_PRL} are also understood by assuming only one hole-like Fermi pocket in the nematic state of FeSe. The picture of one hole-like Fermi surface in FeSe is theoretically understood by considering the nematicity and$/$or spin-orbit coupling\cite{Sprau_Science,DFLiu_PRX2018,Hu_PRB2018,Kang_PRL2018}. Understanding of the physical properties\cite{GYChen_PRB,PKBiswas_PRB,FHardy_PRB} and the pairing mechanism of FeSe \cite{Kreisel_PRB,Benfatto_NPJ} have been based on one Fermi pocket picture around the BZ center in the nematic state.

In this paper, we performed high resolution laser-based ARPES measurements on the electronic structure of FeSe in the nematic state. Double hole-like Fermi surface sheets around the BZ center, and the splitting of the associated bands, are observed for the first time in FeSe. These two Fermi surface sheets exhibit different orbital characters and can not be attributed to the $k_{z}$ effect. These results indicate that  there is an additional order in FeSe, in addition to the nematic order and spin-orbital coupling, that further breaks inversion or time reversal symmetries to give rise to the band splitting. The new Fermi surface picture of FeSe asks for reexamination of the previous theoretical and experimental understandings and stimulates further efforts to identify the origin of the hidden order in the nematic state.

%Figure1

The ARPES measurements were performed by using our latest generation laser-based system; the present observations were made possible because of its unique capabilities of simultaneous coverage of two-dimensional momentum, versatile laser light polarizations and ultra high energy and momentum resolutions (see Methods)\cite{XJZ_Review}. Fig. 1 shows the Fermi surface and constant energy contours of FeSe measured at low temperature in its nematic state under different polarization geometries (Fig. 1a-1c). The crystal structure of FeSe transforms from tetragonal to orthorhombic below 90 K; the distance between adjacent Fe becomes slightly different and $a_{Fe}(b_{Fe})$ is defined as the long $x$ (short $y$) axis in Fig. 1d. Although we did not detwin our FeSe samples intentionally, we find that, in some cases, we can measure on FeSe samples or areas which are dominated by a single domain. This may be caused by local strain that is present or exerted during the sample preparation process and similar phenomenon was also observed before\cite{DFLiu_PRX2018,THashimoto_NC}. Fig. 1a-1c shows results obtained from single domain FeSe. When the electric field vector of the incident laser light, E, is aligned parallel to the $a$ axis (named as $PA$ polarization), the measured Fermi surface mapping (leftmost panel in Fig. 1a) shows an ellipsoid-like shape with the strong spectral weight mainly concentrated on the two vertex areas along the long axis ($b$ direction) and much weaker spectral weight along the short axis ($a$ direction). Upon increasing the binding energy, the corresponding constant energy contours increase in the overall area, consistent with the hole-like nature of the observed Fermi surface. While the weak feature along the short axis remains and the distance between the two parallel sheets slightly increases with increasing binding energy, the two vertex areas along the long axis appear to split and consist of two components. When the electric field vector of the laser light is switched to be parallel to the $b$ direction (named as $PB$ polarization), the Fermi surface mapping in Fig. 1b is dramatically different from that in Fig. 1a. The spectral weight near the two vertex areas along the long axis is strongly suppressed and the observed Fermi surface size along the $a$ axis is obviously larger than that observed in Fig. 1a. By setting the electric field vector of the laser light away from perfect $b$ axis and making use of the combined matrix element effects in Fig. 1a and 1b, we can directly observe two hole-like Fermi surface sheets in Fig. 1c. Combining all the measurements in Fig. 1a, 1b and 1c, we have provided direct evidence that there are two hole-like Fermi surface sheets in FeSe around the BZ center, as depicted in Fig. 1e. This is the first time that two hole-like Fermi pockets around the zone center are clearly observed simultaneously in single domain FeSe.

The orbital character of the observed Fermi surface sheets can also be determined by the polarization measurements, taking advantage of the photoemission matrix element effects. In the $PA$ polarization geometry, for both the horizontal and vertical cuts, only the $d_{xz}$ orbital is allowed\cite{DFLiu_PRX2018,THashimoto_NC,Rhodes_PRB}. On the other hand, in the $PB$ polarization geometry, the $d_{yz}$ orbital is allowed for both the momentum cuts. Therefore, from the spectral weight distribution along the two Fermi surface sheets measured in the $PA$  (leftmost panel in Fig. 1a) and $PB$ (leftmost panel in Fig. 1b) polarization geometries, we have determined the orbital character along the two Fermi surface sheets as shown in Fig. 1e. The entire inner Fermi surface is dominated by the $d_{xz}$ orbital. For the outer Fermi surface sheet, the vertex area along the long axis is dominated by the $d_{xz}$ orbital while the area along the short axis is composed predominantly of the $d_{yz}$ orbital.

%Figure2
The observation of two Fermi surface around the zone center in single domain FeSe can provide a unified picture to understand the Fermi surface topology measured in the twinned samples. Fig. 2 shows the Fermi surface mapping of a twinned FeSe sample measured in $PA$ (Fig. 2d), $PB$ (Fig. 2e) and a mixed (Fig. 2f) polarization geometries. For a direct comparison, we also re-plot the Fermi surface of single domain FeSe in Fig. 2a, Fig. 2b and Fig. 2c measured under the same polarization geometries. The polarization geometry is schematically shown in Fig. 2m and the two kinds of domains are shown in Fig. 2n. Here the axes in the lab coordinates are defined as X, Y and Z while the axes x and y are associated with the particular single domain FeSe.  Under a given polarization geometry, the Fermi surface mappings of the twinned sample (Fig. 2d, 2e and 2f) contain extra features that are not present in the single domain sample (Fig. 2a, 2b and 2c). This can be well understood as the overlap of signals from two orthogonal domains. In the twinned FeSe sample, there are two types of domains, named as vertical domain and horizontal domain that are orthogonal to each other (Fig. 2n). From the single domain data in Fig. 2a-2c, by considering the two orthogonal domains, we can simulate results for the twinned sample as shown in Fig. 2g-2i. For example, under the $PA$ polarization geometry, Fig. 2g is obtained by summing up the signal of the vertical domain in Fig. 2a and that of the horizontal domain in Fig. 2b but rotated by 90 degrees. The simulated results (Fig. 2g-2i) can well reproduce the measured results (Fig. 2d-2f). In this case, the Fermi surface mappings of the twinned sample measured in different polarization geometries (Fig. 2d, 2e and 2f) can be well understood from summation of signals from two types of orthogonal domains, as schematically shown in Fig. 2j, 2k and 2l. The simultaneous observation of four Fermi surface sheets in the twinned sample (Fig. 2f) provides unambiguous evidence on the double Fermi surface around the zone center in the single domain FeSe (Fig. 2c).

The polarization measurements on twinned samples also provide a good way to determine the orbital character on the Fermi surface in a quantitative manner\cite{Rhodes_PRB}. This is realized by comparing the measured spectral weight on the Fermi surface of one horizontal domain and its equivalent Fermi surface of another vertical domain. In Fig. 2e, the measured Fermi surface mapping consists of signals from the horizontal domain (solid lines) and vertical domain (dashed lines), also as schematically shown in Fig. 2k. P1 and P2 are the vertex points of the inner and outer Fermi surface for the horizontal domain while P1' and P2' are the equivalent points on the Fermi surface of another vertical domain. Under the $LV$ polarization (Fig. 2e), according to the matrix element analysis (Fig. 2o), only $d_{xz}$ orbital is allowed for the P1 and P2 points while the $d_{yz}$ orbital is fully forbidden for the horizontal domain. On the other hand, for P1' and P2' points for the vertical domain, the $d_{yz}$ orbital is allowed while the $d_{xz}$ orbital is forbidden under the same $LV$ polarization geometry  (Fig. 2o). Therefore, the $d_{xz}$ and $d_{yz}$ orbital components for the given momentum point, P1 or P2, can be determined from the spectral weight at the two equivalent points from the two orthogonal twins\cite{Rhodes_PRB}.  Fig. 2p shows the measured photoemission spectra (energy distribution curves, EDCs) for P1 and P1' points marked in Fig. 2e, and the EDCs for P2 and P2' points are shown in Fig. 2q. From the EDC peak area, we estimated that for the P1 point, it consists of $\sim$88$\%$ $d_{xz}$ orbital and $\sim$12$\%$ $d_{yz}$ orbital, while for the P2 point it consists of $\sim$84$\%$ $d_{xz}$ orbital and $\sim$16$\%$ $d_{yz}$ orbital. These results confirm that the vertex area of the vertical long axis is dominated by the $d_{xz}$ orbital (Fig. 1e).

%Figure3
Figure 3 shows band structure of FeSe measured along two high symmetry directions under different polarization geometries. Different polarization measurements help provide a complete band structure because the band may show different intensities under different polarization geometries due to the photoemission matrix element effects. The polarization measurements can also help identify the orbital character of the measured bands. To see the measured bands more clearly, we also show the second derivative images (right panels in Fig. 3a-3f) to enhance the contrast, in addition to the original data (left panels in Fig. 3a-3f). Fig. 3g and 3h summarizes the observed bands along the vertical $k_{y}$ and the horizontal $k_{x}$ directions, obtained from the data in Fig. 3a-3c and Fig. 3d-3f,  respectively. Overall, in the measured energy range, three branches of bands (labeled as $\alpha$, $\beta$ and $\gamma$ in Fig. 3g and 3h) are observed for both the momentum cuts, and each branch may be further split into two sub-branch bands. For the bands of the vertical cut (Fig. 3g), the $\alpha_{V1}$ and $\alpha_{V2}$ bands are clearly observed in the raw data (Fig. 3c); the $\beta_{V1}$ and $\beta_{V2}$ bands can be identified in the second derivative image (Fig. 3b) and the weak $\gamma_{V}$ band can be seen in all the second derivative images (Fig. 3a-3c). For the bands of the horizontal cut (Fig. 3h), the $\alpha_{H1}$ and $\alpha_{H2}$ bands are clearly observed in the raw data (Fig. 3f); the $\beta_{H1}$ and $\beta_{H2}$ bands can be identified in the second derivative image (Fig. 3e) and the weak $\gamma_{H1}$ and $\gamma_{H2}$ bands can be seen more clearly in the second derivative image (Fig. 3e).

Our results provide direct evidence on the band splitting and the coexistence of two hole-like Fermi pockets around the BZ center. These observations can not be explained by the $k_{z}$ effect. In the photoemission process, the relatively short photoelectron escape depth gives rise to a finite momentum resolution of $k_{z}$\cite{Hufner}. In FeSe, the $k_{z}$ dispersion of the bands near the zone center is clearly observed at different photon energies between 20$\sim$50 eV\cite{Watson_PRB}. For the 6.994 eV laser we used, it is expected to have an enhanced bulk sensitivity and better $k_{z}$ momentum resolution\cite{GDLiu_RSI}. Therefore, the observation of two separate Fermi pockets can not be due to finite $k_{z}$ resolution because the summation over a range of $k_{z}$ would give a continuous spectrum rather than two discrete features we observed. The 6.994 eV laser we used corresponds to a $k_{z}$ of 0.47 $\pi/c$ that lies nearly in the middle of the basal $k_{z}$=0 and top $k_{z}$=1 planes\cite{Watson_PRB}. We note that signatures of the two hole-like Fermi pockets and the associated band splitting were present in some previous ARPES, STM and quantum oscillation measurements although only one Fermi surface was considered\cite{Watson_PRB,LFanfarillo_PRB,Watson_JPSJ,THashimoto_NC,Kasahara_PNAS,Terashima_PRB,Watson_PRL}.

Three branches of bands have been observed in FeSe around the zone center by previous ARPES measurements\cite{JMaletz_PRB,KNakayama_PRL,TShimojima_PRB,Watson_PRB,Watson2_PRB,Watson_NJP,Watson_JPSJ,PZhang_PRB,Suzuki_PRB,Xu_PRL,DFLiu_PRX2018,Rhodes_PRB,THashimoto_NC,Kushnirenko_PRB,Pfau_PRL}. The polarization measurements combined with the band structure calculations have indicated that the observed three branches of bands are mainly derived from the $d_{xz}$, $d_{yz}$ and $d_{xy}$ orbitals, as schematically shown in Fig. 3i when the nematicity is included without considering the hybridization between bands. From the measured band structure in Fig. 3g and 3h, it is clear that band hybridization occurs at all the band crossing points in Fig. 3j. The resultant band structure by considering these band hybridizations in Fig. 3j shows a good agreement with the measured results (Fig. 3g and 3h) without considering the band splitting. However, even though the nematicity, band hybridization and spin-orbit coupling are all considered, only three bands in the measured energy range are expected (Fig. 3i and 3j).  It is impossible to explain our observation of up to 6 bands in Fig. 3g and 3h. Only when a new mechanism is invoked that can cause further band splitting of the three bands ($\alpha$, $\beta$ and $\gamma$) can we get a band picture (Fig. 3l) that is possibly consistent with the measured results (Fig. 3g and 3h).

%Figure4
The observations of double Fermi pockets and band splitting in FeSe indicate that an additional mechanism is at play in addition to the spin-orbit coupling and the nematic order. The spin-orbit coupling can only lift the degeneracy of $d_{xz}/d_{yz}$ bands at the BZ center. The four-fold rotational symmetry breaking from the nematicity can further enhance the splitting between the $d_{xz}$ and $d_{yz}$ bands. But neither of them can cause the splitting of the $d_{xz}$ or $d_{yz}$ band. Therefore, there should be an additional symmetry breaking that occurs in the nematic phase. We have considered two possible candidates to carry out theoretical simulations (see Methods) and compare them with our measurements. One is the time reversal symmetry breaking (ferromagnetic order) and the other is the inversion symmetry breaking (Rashba spin-orbit coupling), which may originate from the asymmetrical orbital order on top and bottom Se sites or two Fe sites. Fig. 4 shows the measured Fermi surface and the detailed momentum dependence of band structure for FeSe measured under different polarization geometries. The calculated Fermi surface and band structure are shown in Fig. 4g-4h and Fig. 4i-4j, for the ferromagnetic ordering and the Rashba spin splitting effects, respectively. In the case with a ferromagnetic order (Fig. 4i), the prominent features are that the $d_{xz}$, $d_{yz}$ and $d_{xy}$ bands are split and the splitting for the $d_{xz}/d_{yz}$ bands reaches the maximum at $\Gamma$ point and decreases away from $\Gamma$ point. In the latter case, the inversion symmetry breaking introduces Rashba spin-orbit coupling. In contrast to the ferromagnetic ordering, the splitting for the $d_{xz}/d_{yz}$ bands vanishes at $\Gamma$ point and increases away from $\Gamma$ point. It can be found that in terms of current experimental precision, both theoretical calculation results capture well our ARPES measurement results from the Fermi surface topology, the orbital nature evolution and band dispersions along the Fermi surface angles $\theta$. In particular, the double Fermi surface around $\Gamma$ and their orbital nature are both consistent with our experimental results shown in Fig. 1e.

The observed two pockets and band splitting in the nematic phase of FeSe indicate there is an additional order that breaks either the time reversal symmetry or the inversion symmetry. They provide important implications in understanding the superconducting pairing in FeSe. For the ferromagnetic ordering scenario, as each pocket has the same spin, the spin triplet $p$-wave pairing state is likely to be dominant which is consistent with the measured gap function\cite{DFLiu_PRX2018,THashimoto_NC}. Signatures of the time reversal symmetry breaking were reported at the twin boundaries of FeSe in the superconducting state\cite{TWatashige}. For the Rashba scenario, due to the inversion symmetry breaking, spin singlet and spin triplet pairing in principle can mix and an exotic pairing state may occur. Moreover, the two pockets have the opposite spin texture, a sign change between the gap functions on those pockets can render FeSe in a topologically nontrivial phase\cite{Qi2010}, which can host Majorana edge modes. Further measurements are needed to distinguish these two scenarios, in particular, spin resolved experiments can provide a verdict because the Fermi surfaces show distinct spin textures between these two cases.

In summary, by carrying out high-resolution laser-based ARPES measurements on FeSe, double hole-like Fermi surface and band splitting are clearly observed for the first time around the BZ center in the nematic phase of FeSe. The new Fermi surface topology asks for reexamination of the previous theoretical and experimental understanding of FeSe. These results indicate that, in addition to the nematic order and spin-orbit coupling, there is an extra hidden order in FeSe that breaks either inversion or time reversal symmetries. To the best of our knowledge, no candidate of such a new order has been identified experimentally. We hope our present work can stimulate further effort to identify the hidden order and provide new information in understanding the nematicity and superconductivity in FeSe.\\

\noindent {\bf Methods}\\
\noindent{\bf Sample} High quality FeSe single crystals were grown by KCl/AlCl$_{3}$ chemical vapor transport technique\cite{Boehmer_PRB2013}. The samples measured here have a T$_{c}$ of 9.2 K with a transition width of $\sim$0.2 K.\\

\noindent{\bf ARPES Measurements} High-resolution ARPES measurements were performed at our newly developed laser-based system equipped with the 6.994 eV vacuum ultraviolet laser and the time-of-flight electron energy analyser (ARToF 10k by Scienta Omicron)\cite{YZhang_NC,XJZ_Review,DFLiu_PRX2018}. This latest-generation ARPES system is capable of measuring photoelectrons covering two-dimensional momentum space ($k_x$, $k_y$) simultaneously. The system is equipped with an ultra-low temperature cryostat which can cool the sample to a low temperature of 1.6 K. The ARPES measurements were carried out using various polarizations of the laser light. The laser spot size on the sample is about 70 $\mu$m. The overall energy resolution combining the laser linewidth and the ARToF electron energy analyser is $\sim$1 meV. The angular resolution of the analyser is $\sim$0.1 degree, corresponding momentum resolution of 0.0015 $\AA^{-1}$ at a photon energy of 6.994eV. The Fermi level is referenced by measuring polycrystalline gold, which is in good electrical contact with the sample, as well as the normal state measurement of the sample above T$_{c}$. The samples were cleaved {\it in situ} and measured at different temperatures in ultrahigh vacuum with a base pressure better than 5.0$\times$10$^{-11}$ mbar. \\

\noindent{\bf Electronic Structure Simulations} To simulate the band structure of FeSe, we adopted 5-orbital tight-binding model\cite{Su_JPCM,Wu2016,Li_JPCM} including on-site spin-orbit coupling ($\lambda$). In the nematic state, to explain the observed band splitting around $\Gamma$ point, we further consider an $s$-wave nematic order ($\Delta_{s}$) which lifts the degeneracy of $d_{xz}/d_{yz}$ bands and breaks the $C_4$ rotational symmetry. The Hamiltonian of $s$-wave orders are given by
\begin{equation}
H_{s}=\sum_{\boldsymbol{k}}\Delta_{s}(cosk_{x}+cosk_{y})(n_{xz,\boldsymbol{k}}-n_{yz,\boldsymbol{k}})
\end{equation}
where $n_{\alpha,\boldsymbol{k}}=n_{\alpha,\boldsymbol{k}\uparrow}+n_{\alpha,\boldsymbol{k}\downarrow}$ is the density for $\alpha$ orbital. To model the observed band splitting in ARPES experiments, we consider two possible origins: ferromagnetic ordering and Rashba spin-orbit coupling. The ferromagnetic Hamiltonian term is given by
\begin{eqnarray}
H_{FM}=\sum_{\bm{k}\alpha\sigma_1\sigma_2}\bm{M}\cdot\bm{\sigma}_{\sigma_1\sigma_2} c^\dag_{\alpha\bm{k}\sigma_1}c_{\alpha\bm{k}\sigma_2},
\end{eqnarray}
where $\bm{M}$ is the total magnetization. If the inversion symmetry is broken in the nematic phase, Rashba spin-orbit coupling can occur and the term for $d_{xz}$ and $d_{yz}$ orbitals reads,
\begin{eqnarray}
H_{R}&=&\sum_{\bf{k}}[-2(i\lambda_1sink_x+\lambda_2sink_y)c^{\dag}_{xz,\bf{k}\uparrow}c_{xz,\bf{k}\downarrow}-
2(i\lambda_2 sink_x+\lambda_1 sink_y)c^{\dag}_{yz,\bf{k}\uparrow}c_{yz,\bf{k}\downarrow}\nonumber\\
&&-2\lambda_3(isink_x+sink_y)c^{\dag}_{xy,\bf{k}\uparrow}c_{xy,\bf{k}\downarrow}+h.c.]
\end{eqnarray}
In the nematic phase, the adopted parameters in the main text are: $\lambda=10$ meV and $\Delta_s=8.75$ meV. For the scenario of ferromagnetic order induced splitting, the magnetization is assumed along $z$ direction and $M_z=4$ meV is adopted. For the scenario of Rashba spin-orbit coupling induced splitting, $\lambda_1=\lambda_2=10$ meV and $\lambda_3=5$ meV are used.\\

\vspace{3mm}

\noindent {\bf Acknowledgement}\\
We thank financial support from the National Key Research and Development Program of China (Grant No. 2016YFA0300300, 2017YFA0302900, 2018YFA0704200 and 2019YFA0308000), the National Natural Science Foundation of China (Grant No. 11888101, 11874405), the Strategic Priority Research Program (B) of the Chinese Academy of Sciences (Grant No. XDB25000000), the Youth Innovation Promotion Association of CAS (Grant No.2017013), and the Research Program of Beijing Academy of Quantum Information Sciences (Grant No. Y18G06).

\vspace{3mm}

\noindent {\bf Author Contributions}\\
 C.L., X.X.W., and L.W. contributed equally to this work. X.J.Z. and C.L. proposed and designed the research. L.W. and Y.G.S. contributed to FeSe crystal growth. X.X.W., C.X.D., J.P.H. and T.X. contributed to the band structure calculations and theoretical discussion. C.L., D.F.L., Y.Q.C., Y.W., Q.G., C.Y.S., J.W.H., J.L., P.A., G.D.L., Y.H., Q.Y.W., F.F.Z., S.J.Z., F.Y., Z.M.W., Q.J.P., Z.Y.X., L.Z. and X.J.Z. contributed to the development and maintenance of Laser-ARTOF system. C.L., Q.G. and X.W.J. contributed to software development for data analysis. C.L., Q.Y.C. and Y.W. carried out the experiment with the assistance from Q.G., C.Y.S., J.W.H., J.L., P.A., H.L.L. and C.H.Y.. C.L., D.F.L., Z.L. and X.J.Z. analyzed the data. C.L., Z.L. and X.J.Z. wrote the paper with X.X.W., J.P.H. and T.X.. All authors participated in discussion and comment on the paper.

\newpage

\begin{figure*}[tbp]
\begin{center}
\includegraphics[width=1\columnwidth,angle=0]{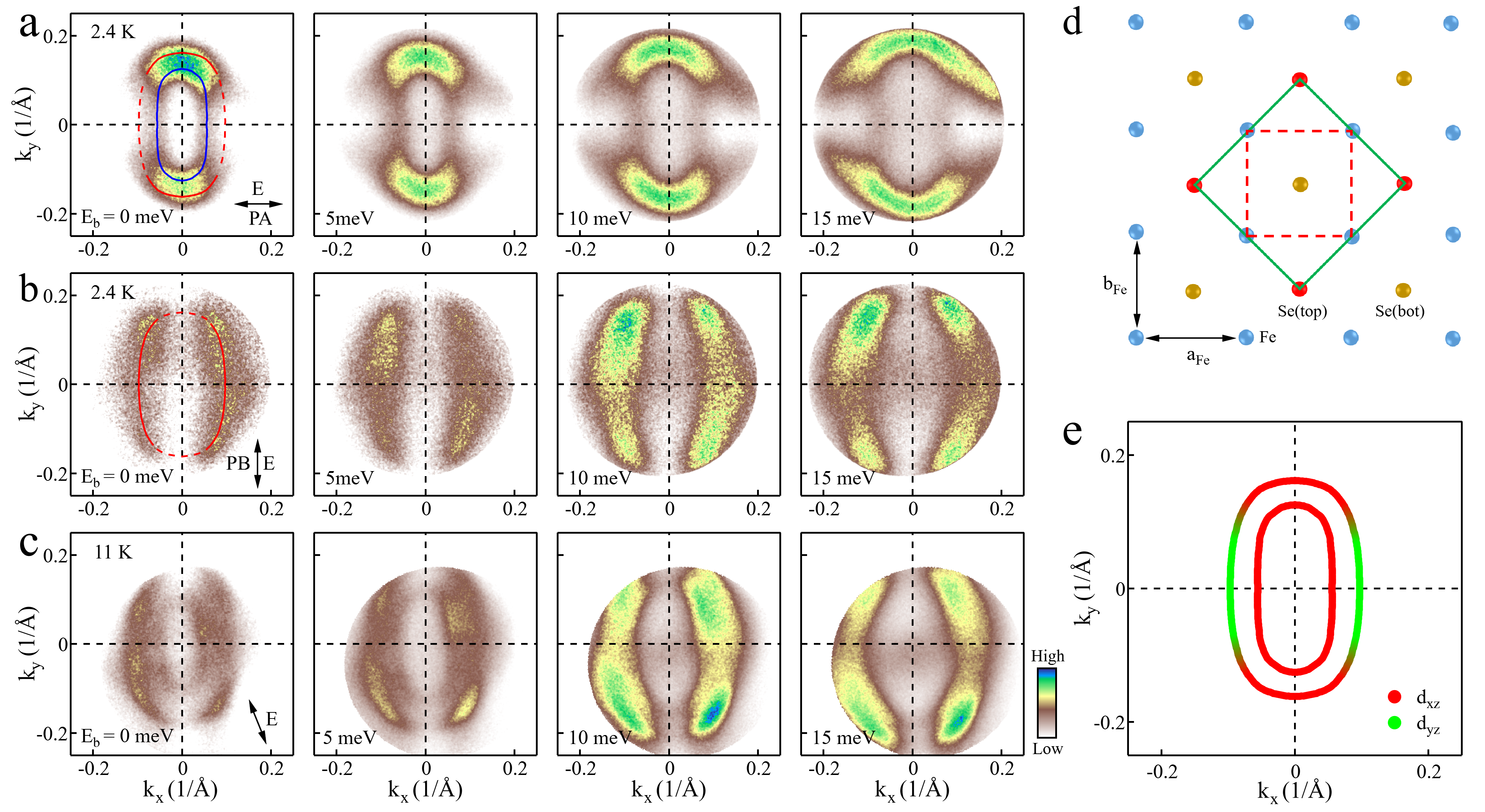}
\end{center}
\caption{\textbf{Observation of two hole-like Fermi surface sheets in single domain FeSe.} (a-c) Constant energy contours of the single domain FeSe at 2.4 K under the $PA$ polarization (a), at 2.4 K under the $PB$ polarization (b) and at 11 K under a mixed polarization (c). The direction of the electric field vector E corresponding to the three polarization geometries is marked by double arrows in the leftmost panels in (a-c). Blue and red ellipses in the leftmost panels are the guide lines of the observed Fermi surface. (d) Top view of FeSe crystal structure. The blue, red and yellow circles represent Fe atom, Se atom above the Fe plane (Se top) and Se atom below the plane (Se bottom), respectively. The red dashed line represents the 1-Fe unit cell while the green solid line represents 2-Fe unit cell. We define the $x$ axis parallel to the $a_{Fe}$ axis and the $y$ axis parallel to the $b_{Fe}$ axis. In the normal state, FeSe has $C_{4}$ symmetry with $a_{Fe}$=$b_{Fe}$, but it breaks into $C_{2}$ symmetry in the nematic state with $a_{Fe}$$>$$b_{Fe}$. (e) The extracted Fermi surface of FeSe which consists of two hole-like Fermi surface sheets around the BZ center. The dominant orbital character of the two Fermi surfaces is also marked.
}
\end{figure*}

\begin{figure*}[tbp]
\begin{center}
\includegraphics[width=1\columnwidth,angle=0]{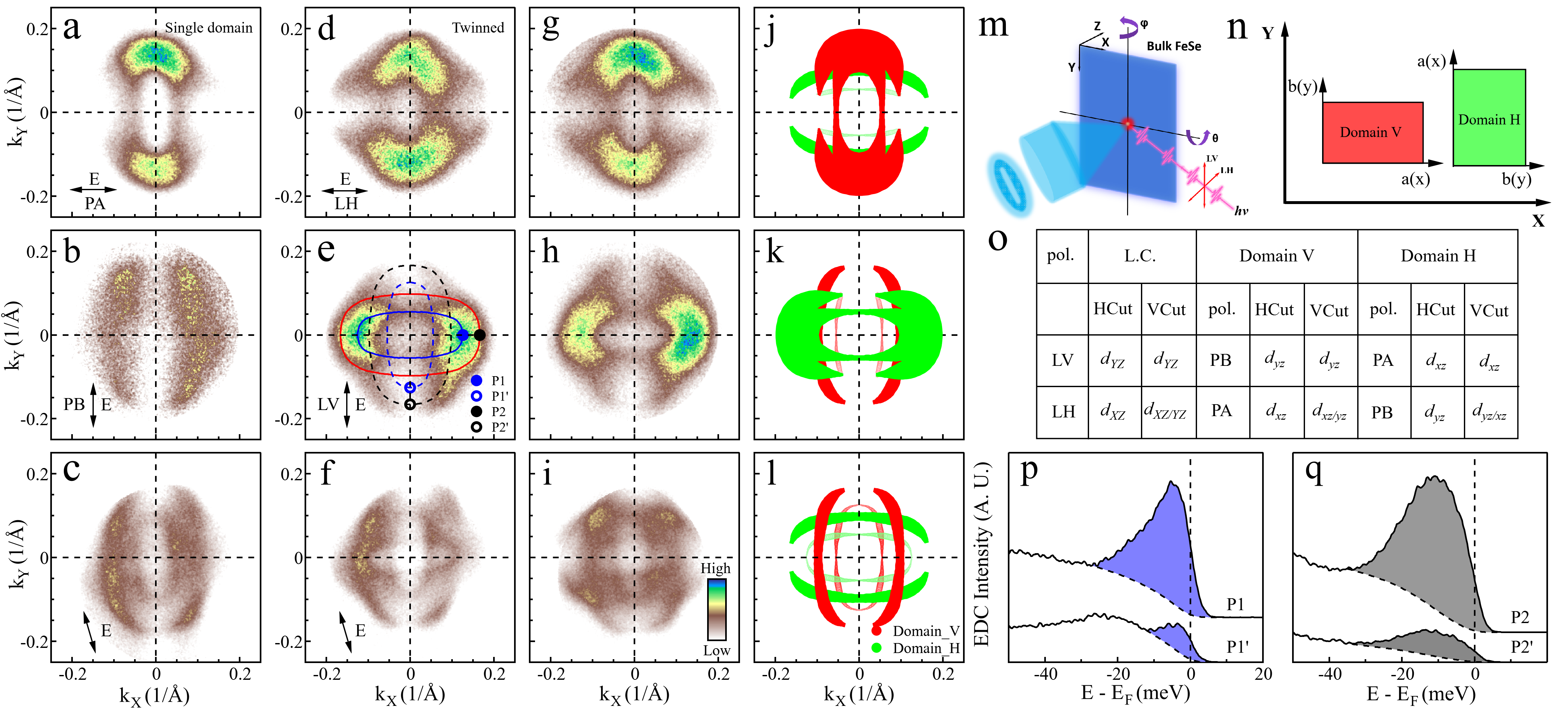}
\end{center}
\caption{\textbf{Fermi surface mapping of twinned FeSe.} (a-c) The Fermi surface mappings of the single domain FeSe measured under the $PA$ (a), the $PB$ (b) and a mixed (c) polarization geometries. (d-f) The Fermi surface mappings of the twinned FeSe measured under the $LH$ (d), the $LV$ (e) and a mixed (f) polarization geometries. (g-i) Simulated Fermi surface from the single domain data in (a-c). Under the $PA$ polarization geometry, (g) is obtained by summing up the signal of the vertical domain in (a) and that of the horizontal domain in (b) but rotated by 90 degrees. Under the $PB$ polarization geometry, (h) is obtained by rotating the data in (g) by 90 degrees. Under the mixed polarization geometry, (i) is obtained by the summation of signals from the vertical domain in (c) and from the horizontal domain also in (c) but rotated 90 degrees. (j-l) Schematic of the observed Fermi surface in twinned FeSe from the horizontal domain (green lines) and the vertical domain (red lines) corresponding to data in (d-f). (m) Schematic drawing of the experimental geometry with $LV$ and $LH$ polarizations. X, Y and Z axes are defined as laboratory coordinates (L.C.). (n) Schematic drawing of two domains in nematic state in lab coordinates. The two domains are rotated 90 degree with each other. For domain V, the $x$ ($y$) axis is parallel to X(Y) axis in L.C. while for domain H, it is reversed. (o) Summary of the allowed orbitals under $LV$ and $LH$ polarization geometries defined for lab coordinates and each domains along the high symmetry cuts.  (p-q) The photoemission spectra (EDCs) on the Fermi surface of two domains. The location of the four momenta is marked in (e). For each EDC, the peak area is obtained after subtracting the background. The obtained area ratio between the P1 and P1' points is 7.3 in (p) and between the P2 and P2' points is 5.3 in (q). The ratio is used to estimate the orbital components of $d_{xz}$ and $d_{yz}$ for the measured momentum points.
}
\end{figure*}

\begin{figure*}[tbp]
\begin{center}
\includegraphics[width=1\columnwidth,angle=0]{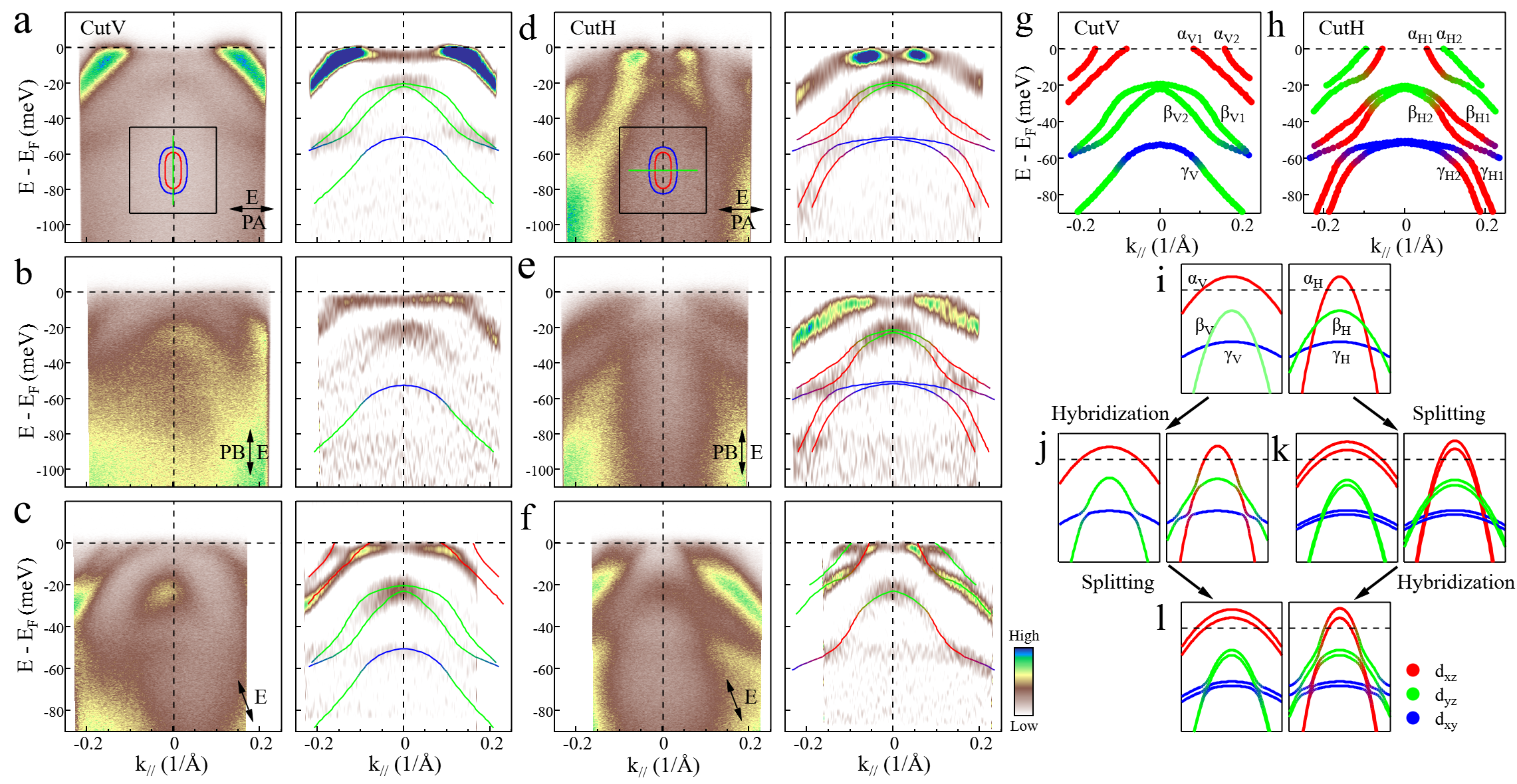}
\end{center}
\caption{\textbf{Band structure of single domain FeSe along high symmetry cuts and their orbital nature.} (a-c) Band structure measured along the vertical momentum cut under $PA$ (a), $PB$ (b) and a mixed (c) polarization geometries. The location of the momentum cut is shown in the inset of (a).  The left panels of (a-c) represent the original data while the right panels are corresponding second derivative images with respect to energy in order to enhance the contrast. Some observed bands are marked by the guide lines. (d-f) Same as (a-c) but for the horizontal momentum cut as marked in the inset of (d). (g-h) Extracted band structure for the vertical momentum cut (g) and horizontal momentum cut (h) obtained from (a-c) and (d-f), respectively. The observed bands are labeled as three branches of $\alpha$, $\beta$ and $\gamma$. (i) Schematic of the band structure along the vertical (left panel) and horizontal (right panel) momentum cuts from $\alpha$ (red lines), $\beta$ (green lines) and $\gamma$ (blue lines) orbitals. (j) Same as (i) but only considering the hybridization between different bands. (k) Same as (i) but only considering the splitting of each band. (l) Same as (i) but considering both band hybridization and splitting of each band. The orbital character of the $d_{xz}$, $d_{yz}$ and $d_{xy}$ orbitals is marked by lines with red, green and blue colors, respectively.
}
\end{figure*}

\begin{figure*}[tbp]
\begin{center}
\includegraphics[width=1\columnwidth,angle=0]{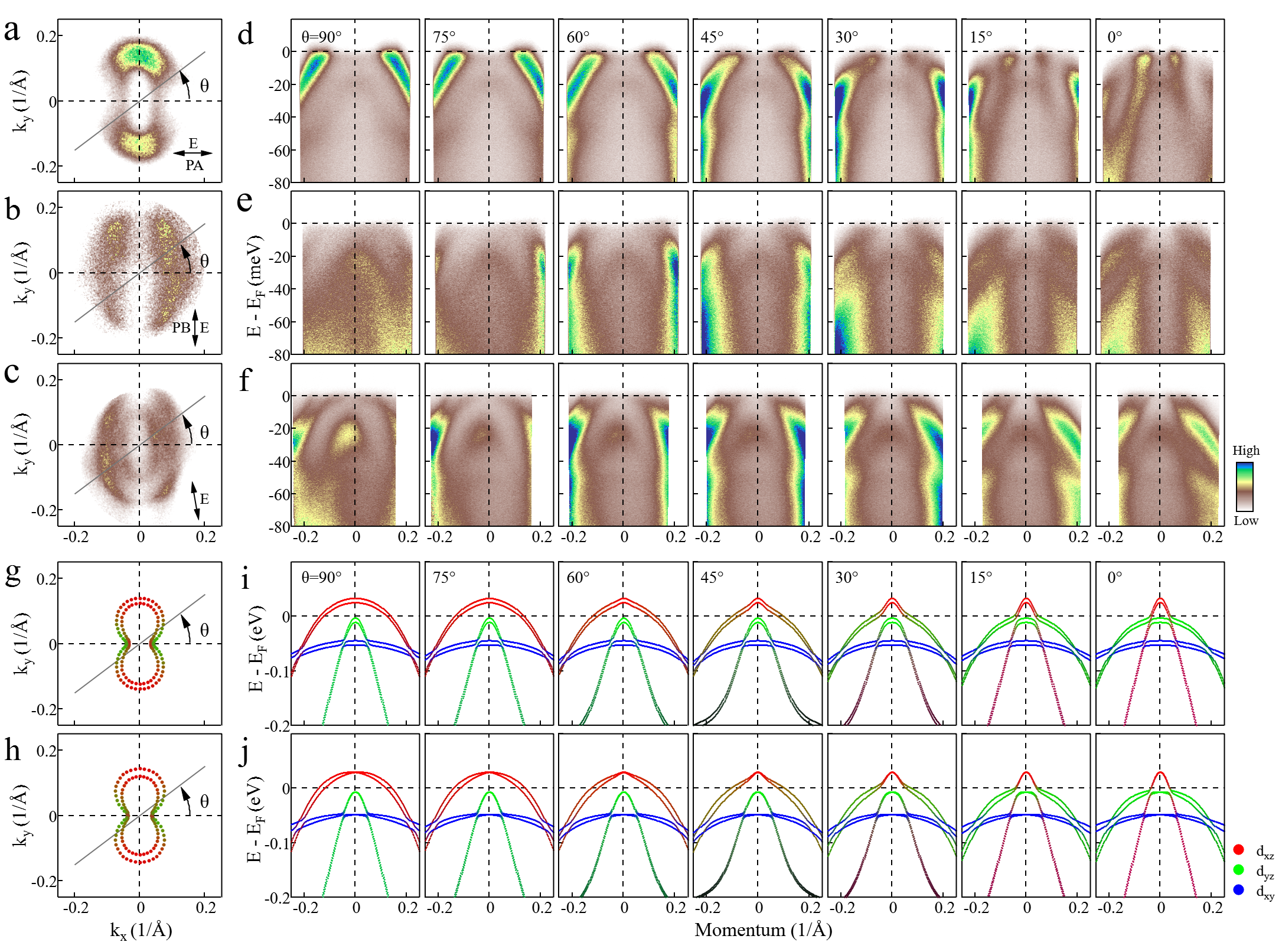}
\end{center}
\caption{\textbf{Calculated Fermi surface and band structure of FeSe and their comparison with the measured results.} (a-c) The Fermi surface mappings of the single domain FeSe measured under the $PA$ (a), the $PB$ (b) and a mixed (c) polarization geometries. The momentum cuts here are defined by the Fermi surface angle $\theta$. (d-f) Band structure for the momentum cuts at different Fermi surface angles $\theta$ corresponding to (a-c). (g-h) Calculated Fermi surface around $\Gamma$ point in considering the ferromagnetic order and the Rashba spin-orbit coupling in the nematic state of FeSe, respectively. Their corresponding band structures are shown in (i-j). The $d_{xz}$, $d_{yz}$ and $d_{xy}$ orbital components are represented by red, green, and blue colors, respectively.
}
\end{figure*}

\end{document}